\documentclass[manuscript]{acmart}
\AtBeginDocument{%
  \providecommand\BibTeX{{%
    \normalfont B\kern-0.5em{\scshape i\kern-0.25em b}\kern-0.8em\TeX}}}

\setcopyright{acmcopyright}
\copyrightyear{2023}
\acmYear{2023}
\acmDOI{}

\acmConference[FashionXRecSys '23]{Fifth Workshop on Recommender Systems in Fashion}{September 18--22,
  2023}{Singapore}
\acmBooktitle{FashionXRecSys'23: Workshop on Recommender Systems in Fashion,
September 18, 2023, Singapore} 
\acmISBN{}

\newcommand{\myparagraph}[1]{\vspace{5pt}\noindent{\bf #1}}
\newcommand{\Ftwoscore}{$\mathrm{F_{2}}$-score}

\usepackage{cleveref}
\usepackage{hyperref}

\begin{document}

%%
%% The "title" command has an optional parameter,
%% allowing the author to define a "short title" to be used in page headers.
\title{UNICON: A unified framework for behavior-based consumer segmentation in e-commerce}

%%
%% The "author" command and its associated commands are used to define
%% the authors and their affiliations.
%% Of note is the shared affiliation of the first two authors, and the
%% "authornote" and "authornotemark" commands
%% used to denote shared contribution to the research.
\author{Manuel Dibak}
\authornote{Authors contributed equally to this work.}
\email{manuel.dibak@zalando.de}
\author{Vladimir Vlasov}
\authornotemark[1]
\email{vladimir.vlasov@zalando.de}
\author{Nour Karessli}
\authornotemark[1]
\email{nour.karessli@zalando.de}
\author{Darya Dedik}
\email{darya.dedik@zalando.de}
\author{Egor Malykh}
\email{egor.malykh@zalando.de}
\author{Jacek Wasilewski}
\email{jacek.wasilewski@zalando.de}
\author{Ton Torres}
\email{ton.torres@zalando.de}
\author{Ana Peleteiro Ramallo}
\email{ana.peleteiro.ramallo@zalando.de}
\affiliation{
  \institution{Zalando SE}
  \city{Berlin}
  \country{Germany}
}

%%
%% By default, the full list of authors will be used in the page
%% headers. Often, this list is too long, and will overlap
%% other information printed in the page headers. This command allows
%% the author to define a more concise list
%% of authors' names for this purpose.
\renewcommand{\shortauthors}{Dibak and Vlasov, et al.}

%%
%% The abstract is a short summary of the work to be presented in the
%% article.
\begin{abstract}
Data-driven personalization is a key practice in fashion e-commerce, improving the way businesses serve their consumers needs with more relevant content. While hyper-personalization offers highly targeted experiences to each consumer, it requires a significant amount of private data to create an individualized journey. To alleviate this, group-based personalization provides a moderate level of personalization built on broader common preferences of a consumer segment, while still being able to personalize the results. We introduce UNICON, a unified deep learning consumer segmentation framework that leverages rich consumer behavior data to learn long-term latent representations and utilizes them to extract two pivotal types of segmentation catering various personalization use-cases: \textit{lookalike}, expanding a predefined target seed segment with consumers of similar behavior, and \textit{data-driven}, revealing non-obvious consumer segments with similar affinities. We demonstrate through extensive experimentation our framework effectiveness in fashion to identify lookalike Designer audience and data-driven style segments. Furthermore, we present experiments that showcase how segment information can be incorporated in a hybrid recommender system combining hyper and group-based personalization to exploit the advantages of both alternatives and provide improvements on consumer experience.
\end{abstract}

%%
%% The code below is generated by the tool at http://dl.acm.org/ccs.cfm.
%% Please copy and paste the code instead of the example below.
%%


%%
%% Keywords. The author(s) should pick words that accurately describe
%% the work being presented. Separate the keywords with commas.

%% A "teaser" image appears between the author and affiliation
%% information and the body of the document, and typically spans the
%% page.
\begin{teaserfigure}
  \includegraphics[width=\textwidth]{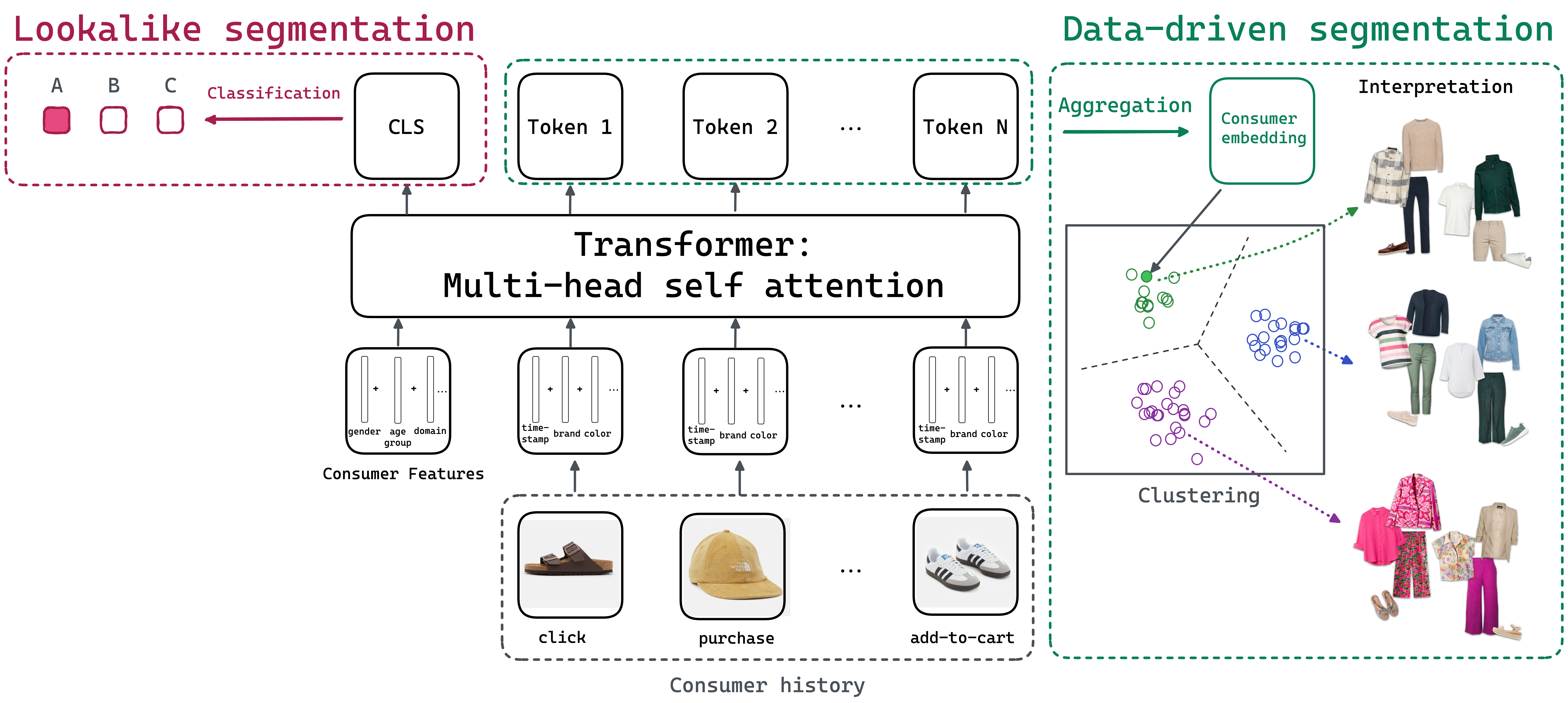}
  \caption{Schematic figure of UNICON: A sequence of consumer interactions is embedded into tokens. A multi-head self attention transformer network generates embeddings
  for these tokens which are then used for lookalike and data-driven consumer segmentation.
  }
  \label{fig:teaser}
\end{teaserfigure}

% \received{20 February 2007}
% \received[revised]{12 March 2009}
% \received[accepted]{5 June 2009}

%%
%% This command processes the author and affiliation and title
%% information and builds the first part of the formatted document.

% \begin{teaserfigure}
%     \includegraphics[width=\textwidth]{figures/example group.png}
%     \caption{figure caption}
%     \label{fig:style-group-schema}
%     \Description{figure description}
% \end{teaserfigure}

\maketitle

\section{Introduction}
\label{sec:introduction}
Personalizing experiences such as search, ranking and recommendation in e-commerce is a fundamental capability to provide consumers with relevant content that is tailored to their specific preferences. Group-based personalization~\cite{perez2021content, dara2020survey, delic2018group, huang2021novel, cao2018attentive, sankar2020groupim, zhang2022gbert,delic2018group}, in comparison to hyper-personalization~\cite{liang2018variational,Choi2006hyper,Song2014hyperranking,bennett2012hypersearch}, allows serving consumers with a customized experience based on broader patterns observed in the grouped segments. This can be particularly beneficial in addressing the cold start problem~\cite{dara2020survey} appearing in new consumers with limited amounts of signals as well as low-engaged consumers whose historical signals can be outdated.
Profiting from collective preferences within a segment, group-based personalization improves the quality of recommendations when individual recommendations are not relevant~\cite{Baltrunas2010} as well as the diversity of the recommended content encouraging consumers to explore new and more diverse items~\cite{Zhang2016}. The quality of group-based personalization heavily depends on the quality of the segments and their representation of the user base. The majority of the reviewed work on group-based personalization assumes consumer segments already exist, based on external social interactions like viewing the same TV program or touristic packages \cite{castro2018fuzzy, de2014comparison}, going to the same venue~\cite{sankar2020groupim, zhang2022gbert} or having the same friends networks~\cite{zhou2021group}. Another common segmentation criterion are ratings of the same movie~\cite{kavsvsak2016personalized, perez2021content, huang2021novel}.

In this work, we develop a method to identify high quality consumer groups that can be used to improve personalized experiences and propose \emph{UNICON - a UNIfied framework for behavior-based CONsumer segmentation in e-commerce}. We distinguish between two important types of consumer segmentation: 1) Lookalike segmentation, where given a predefined seed audience segment (often defined by a business rule), the goal is to expand this segment using ML-based models that use consumer behavior data and predict those that don't (yet) match the business rule but are behaving highly similar to the seed segment~\cite{Ma2016,rahman2023exploring} (e.g., consumers that have never purchased a certain brand but may have affinity towards it), and 2) Data-driven segments, where the groups emerge from the consumer data and allows us to find non-obvious segments of similar consumers that are too complex to be captured with a set of rules \cite{brito2015customer, zhou2021group, peng2023finding}.

Our work main contributions can be summarized as follows: 1) a consumer unified long-term representation using rich consumer behavior data in a multi-headed self attention transformer-based~\cite{transformer2017,celikik2022reusable} model, 2) two methods that utilize the latent embeddings to extract two types of consumer segments, namely, lookalike and data-driven segments and apply on real-world data at scale to define lookalike Designer segment and data-driven style-based segments, and 3) experiments and results for real life use cases with both data segments, one incorporating it in a hybrid recommender system in a product carousel and the other to improve the experience for Designer consumers in the catalog ranking. \Cref{fig:teaser} illustrates the proposed framework.

The remainder of this paper is structured as follows: in
~\autoref{sec:related-work} we discuss the related literature; in ~\autoref{sec:methodology} we detail the proposed UNICON framework and methods; in~\autoref{sec:experiments-and-results} we share offline and online experimental results; finally, in~\autoref{sec:conclusion} we conclude the work and briefly discuss future work. 

\section{Related work}
\label{sec:related-work}
\subsection{Consumer segmentation}
The exploration of lookalike modeling and data-driven segmentation approaches for recommender systems remains an ongoing area of investigation in the scientific literature. In \cite{brito2015customer} authors perform consumer segmentation based on the properties of ordered shirts. In \cite{zhou2021group} authors propose a model that highlights similar users based on their behavior and friend network using neural networks. The users who appear in friend circles are used to build the group profile. Personalized search results are created by combining individual and group profiles with respect to the current query. \cite{peng2023finding} introduces a real-time attention based lookalike model that dynamically captures user preferences and behavior in real-time to identify similar users. By incorporating attention mechanisms, the model can effectively weigh the importance of different user attributes and adapt to changing user preferences. The model is trained using true similarity scores known for the training dataset. In a related line, \cite{rahman2023exploring} presents an innovative method for cross service lookalike modeling to improve user-targeting in campaigns. Authors propose a rule-based associative classification model that identifies meaningful associations between user attributes and conversion likelihood. By leveraging these associations, the model is able to identify users with similar characteristics, leading to enhanced conversion rates. More specifically in the context of fashion, segmentation based on style has been explored with the help of computer vision methods. In \cite{mall2019geostyle} authors detect local periodically recurring fashion trends from images taken at the same city. In \cite{matzen2017streetstyle} images from photo-sharing services and social media platforms were analyzed using computer vision models to find global and per-city fashion choices and spatio-temporal trends. In \cite{kiapour2014hipster} authors designed a “game” to get training data by letting a user compare two images related to their style, e.g. ``Who’s more hipster?'' and use the collected data to build an inter style classifier (``Is this image goth or hipster?'') and intra style ranker (``How hipster is this image?'').

\subsection{Clustering approaches}
In the data-driven segmentation case, no additional information about consumers such as lookalike labels, friend network or true similarity scores is known. Therefore, we leverage unsupervised clustering approaches. Our consumers are represented as a sequence of interactions with clothing articles. The problem of grouping sequences of tokens is reminiscent of the topic modeling problem in NLP. In \cite{angelov2020top2vec, grootendorst2022bertopic} authors propose to use document encoders~\cite{le2014distributed} or sentence sequential encoders~\cite{reimers2019sentence} as embeddings then use clustering algorithms.
Other works \cite{mcinnes2017accelerated, wang2019attributed} propose to learn embeddings and groups jointly using deep clustering methods. In this work, we adapt separate embedding and clustering approaches to leverage an already existing consumer embedding model~\cite{celikik2022reusable}.

\subsection{Group recommendations}
Group recommendation systems \cite{delic2018group, dara2020survey, perez2021content} are mostly related with the problem of generating recommendations to predefined groups of individuals. They distinguish between persistent and ephemeral groups. While the first one can be considered static groups~\cite{huang2021novel, cao2018attentive}, like families, the latter are considered groups that spontaneously form for a specific activity~\cite{sankar2020groupim, zhang2022gbert}, e.g., a group of friends coming together to go to a restaurant. In these works attention mechanisms are used to capture user preferences and model group interactions.

Unlike the previous methods, in this work we provide a unified framework for identifying consumer segments in either supervised or unsupervised manner and leveraging these segments in a hybrid recommender system.

\section{Methodology}
\label{sec:methodology}
We represent consumers as a sequence of interaction (article-click, add-to-cart, add-to-wishlist and checkout) on articles identified by their stock keeping unit (SKU). 
Using a transformer-based model, we can learn the consumer behavior and subsequently find segments of consumers that interact similarly.

We identify two separate ways of defining segments, \emph{Data-driven segmentation} and \emph{Lookalike segmentation}. Both methods can be summarized in the \emph{UNICON} framework, which involves the steps below. \Cref{fig:teaser} illustrates an overview how both types of segmentation arise from the consumer histories.
\begin{enumerate}
    \item Data preparation: This involves selecting the right time frame of interactions and the right attributes of the interacted items to naturally form segments of certain properties. For lookalike segmentation this also involves identifying consumers that fulfill the business logic and can be used as training examples.
    \item Next item prediction training (data-driven only): This trains the transformer to predict the next item (token) the consumer will interact with by using a softmax classifier with cross entropy loss and masking out future interactions. This ensures the model finds an internal representation of the consumer and naturally groups sequences together that would likely interact with the same items and therefore are similar in behavior.
    \item Classification training (lookalike only): This step trains the model to predict membership of the consumer segment based on the business logic. The classification uses the CLS token as input and produces scores for each of the classes through a softmax layer.
    \item Embedding extraction (data-driven only): This aggregates the embeddings of all the interactions of the consumer into a single point in a high-dimensional space in which proximity corresponds to similarity in terms of behavior, meaning that consumers close together will likely interact with the same type of items.
    \item Clustering (data-driven only): In a final step, the groups are found using clustering to identify segments in the embedding space, which correspond to segments of consumers.
\end{enumerate}

\subsection{Data-driven segmentation}
\emph{Data-driven segmentation} aims at learning to partition consumers into segments solely based on their interactions. This approach uses unsupervised learning to group consumers together that exhibit similar behavior patterns. For data driven segmentation, we deploy a two-step approach: 1) we learn consumer embeddings in which consumers exhibiting similar behavior are close to each other by a distance, or similarity metric. Subsequently, 2) we divide this space into groups such that a consumer can belong to only one group. 

\myparagraph{Consumer Embeddings.} We employ a transformer-based model to compute consumer embeddings from a sequence of interactions with items. Each item in the input sequence is represented by various attributes such as brand, color, silhouette, etc. The model is trained to predict the SKU of the next item interaction using a causal self-attention mechanism and a categorical cross-entropy loss~\cite{celikik2022reusable}. The consumer embeddings are extracted from the encoder’s output, by averaging over the embedded token sequence. All interactions are treated equally in the averaging.

\myparagraph{Segmentation.} Once the embedding has been created, the segmentation is performed by applying the k-means clustering algorithm on the consumer embeddings.

\subsection{Lookalike segmentation}
\emph{Lookalike segmentation} on the other hand use a set of consumers defined by business logic and aims at finding consumers that behave similarly to them, i.e., aims at extending groups of consumers defined by business rules. This approach learns the typical behavior of such segments based on their interaction sequences in order to find consumers that behave similarly and would likely fulfill the business rules in the future. 
Formally, let $C$ be the set of consumers fulfilling the business logic. We call this set of consumers the \emph{core segment}. Further, let $C^{E}$ be the set of consumers that in the future will become part of the core segment. We are interested in identifying the set of \emph{lookalike consumers} $L $, which will likely be part of the core segment in the future:
\begin{equation}
L = \left\{ l \mid l \notin C \wedge p\left(l \in C^{E} \right) > \tau \right \},
\label{eq:lookalike-set}
\end{equation}
meaning that their probability of becoming a core consumer in the future is higher than some pre-defined threshold $\tau$. 

This definition of lookalike consumers allows for a formulation of the identification of the \emph{lookalike segment} as a supervised learning problem, where we use the set of core consumers as positive examples, and random non-core consumers as negative examples for a binary classifier based on the user history.

\section{EXPERIMENTS AND RESULTS}
\label{sec:experiments-and-results}
We identified two use cases to apply data-driven and lookalike segmentation. For data-driven segmentation, we aim at identifying segments of consumers that show affinity to a similar fashion style. This is a challenging task, as there is no straightforward way of quantifying a style similarity. As a part of this project, we propose a method to address this problem and to automate the evaluation of the consumer segments.
As an application of lookalike segmentation, we identify consumers that are likely to be interested in designer items. Designer items are those produced by prestigious and luxurious brands. To this end, we define \emph{designer consumers} based on rules applied to consumers' histories and apply the method to find a set of \emph{designer lookalikes}.

\subsection{Data-driven segmentation}
\subsubsection{Data-driven style segmentation}
As we want to find segments of consumers that are similar in style, we need to only present the model with information that is relevant for style identification. We achieve this by filtering the consumer histories to only contain information about interaction that might indicate a certain style affinity of the consumer.

\myparagraph{Data variants.} 
We explore several data variants: \textbf{Baseline)} consumer sequences with interactions (article-click, add-to-wishlist, add-to-cart, and checkout) on any SKUs over the previous two months. These sequences can have mixed SKUs in terms of fashion preference (female, male) (39.1M sequences). \textbf{V1)} filtering consumer sequences to contain only style relevant silhouettes defined by a fashion expert (37M sequences). \textbf{V2)} V1 + splitting consumer sequences by fashion gender preference (26.4M male sequences and 33M female sequences). \textbf{V3)} V1 + removing sequences that contain only a single silhouette type (28.7M sequences). \textbf{V4)} Combine the filters of variants V1 and V3 (16.5M male sequences and 24.1M female sequences).

\myparagraph{Consumer embedding and segmentation offline experiments.} The goal is to validate that proximity in the embedding space reflects similarity in style. Additionally, this allows us to identify the optimal distance metric in the embedding space as the one that correlates most with style similarity. 

As a first step, we define a style similarity score $S$ between two users $u_1$ and $u_2$ as  
\begin{equation}
S(u_1, u_2) =  \sum_{a \in A} w_a \left [1 - \mathrm{JSd}\left (P(a, u_1) || P(a, u_2) \right ) \right ],
\label{eq:style-similarity}
\end{equation}
where $A$ is the set of item attributes (e.g. brand, commodity group, color, etc.). $w_a$ is the weight assigned to the attribute $a$, which is selected to match the behavior on samples of user histories annotated by a fashion expert. $P(a, u)$ is the distribution of item attribute $a$ values in the consumer interaction history $u$. $\mathrm{JSd}(P || Q)$ is the Jensen-Shannon divergence. We use this score to evaluate the embedding space variants and to determine which distance metric to use for clustering. To this end, we compute the Pearson correlation between style similarity and distance metrics in the embedding space based on randomly sampled consumer pairs.  \Cref{tab:emb} reports Pearson correlation of style similarity score, with three distance metrics Dot product, Cosine Similarity and Euclidean distance. We can see that all distance metrics show moderate correlation with the style similarity and representation of style increases when more restrictions are applied to data. We choose Cosine similarity and V4 as they exhibit the highest correlation with style similarity score.

\begin{table}
  \caption{Evaluation of the embedding space: Pearson correlation between style similarity defined by \cref{eq:style-similarity} and different distance/similarity metrics in the consumer embedding space.}
  \label{tab:emb}
  \begin{tabular}{rccccc}
    \toprule
    Distance/similarity metric&Baseline&V1&V2&V3&V4\\
    \midrule
    Dot product ↑ &0.435 & 0.471 & 0.481 & 0.476 & \textbf{0.535}\\
    Cosine similarity ↑ & 0.451 & 0.482 & 0.496 & 0.494 & \textbf{0.545}\\
    Euclidean distance ↓ & -0.398 & -0.409 & -0.351 & \textbf{-0.456} &-0.433\\
    \bottomrule
  \end{tabular}
\end{table}

To measure if the segments are well separated in the embedding space we use Silhouette Score~\cite{rousseeuw1987silhouettes} which measures how similar data points are to their own segment as compared to other segments. The score ranges between -1 and 1, with a higher score indicating a better cluster separation. Furthermore, to analyze whether the segmentation is reasonable in terms of style similarity, we use the area under the receiver operating characteristics curve (ROC-AUC) as a consistency metric between style similarity and the consumer segmentation. We compute ROC-AUC for pairs of consumers where style similarity is used as a score and the label is whether they belong to the same style segment.

\begin{table}
  \caption{Evaluation of the number of segments using k-means clustering.}
  \label{tab:nsegments}
  \begin{tabular}{rccc}
    \toprule
    Number of segments&20&250&1000\\
    \midrule
    Silhouette score (cosine) ↑ & \textbf{0.096} & 0.097 &0.087\\
    ROC-AUC Style Similarity ↑ & 0.77 & 0.88 & \textbf{0.91}\\
    \bottomrule
  \end{tabular}
\end{table}

\Cref{tab:nsegments} shows a comparison between methods with different numbers of segments. It can be seen that ROC-AUC is better for a large number of segments while silhouette score is marginally better for a lower number. To find the right number of segments, we further investigate the distribution of distances to the segment center for points inside the segments.

\begin{figure}[h]
  \centering
  \includegraphics[width=\linewidth]{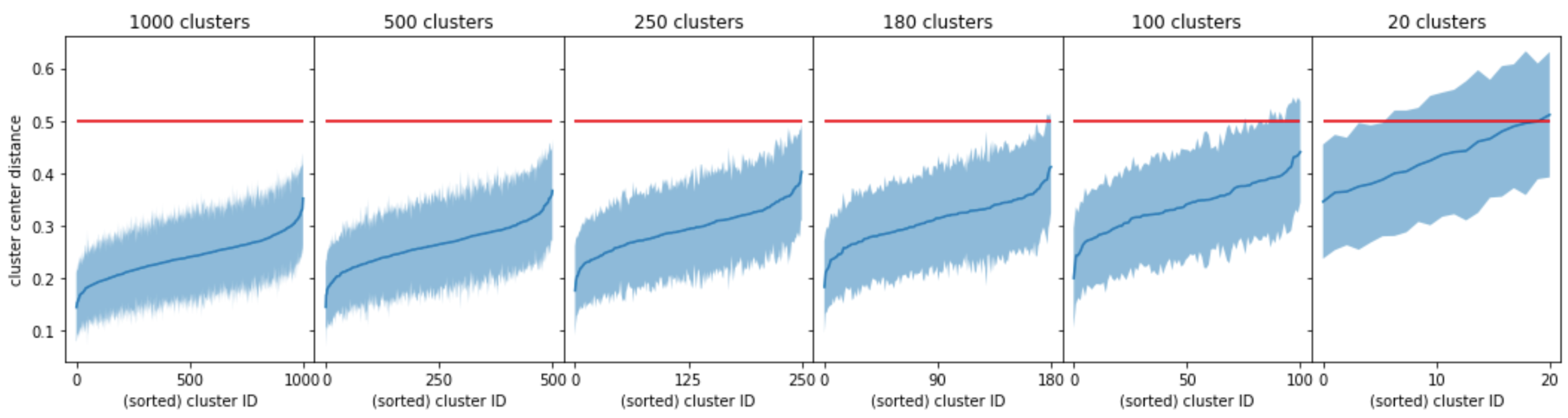}
  \caption{The average and standard deviation of center distances per cluster. The red line indicates the style similarity length scale, which approximates the distance in the embedding space at which the style similarity decays to 1/e. It is determined by fitting an exponential curve to the averaged style similarity over distances binned to a regular grid.}
  \Description{}
  \label{fig:nclusters}
\end{figure}

\Cref{fig:nclusters} shows the average and standard deviation of center distances per segment. The red line indicates the style similarity length scale which approximates the distance in the embedding space at which the style similarity considerably decays. We chose 250 segments because it is a good trade-off between number and size of segments.

\begin{figure}[h]
  \centering
  \includegraphics[width=\linewidth]{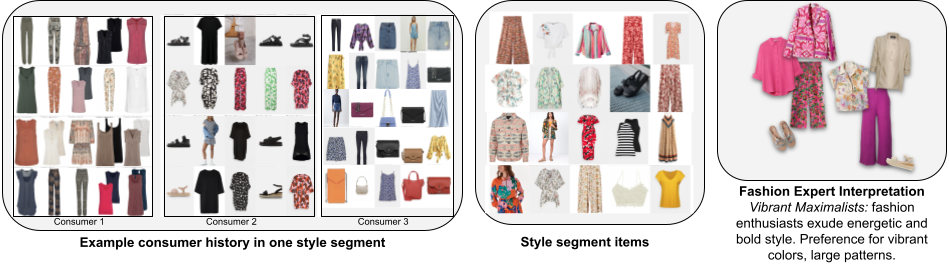}
  \caption{The interaction history of example consumers that were identified to belong to one segment (left), the representative segment items identified in their segment (middle), and the fashion expert interpretation of the style (right).}
  \label{fig:example_segment}
\end{figure}

\myparagraph{Representative segment items}
 We utilize segmentations of consumers to drive a better experience in recommendations. To this end, we make use of \emph{representative segment items}. These are items that represent the consumers in the given segment. To identify representative segment items, we select the top 100 most popular items by significant actions for each segment and each gender preference. 
 To additionally boost diversity in the items and increase their significance to the cluster, we enforce sampling rates based on commodity groups, and only consider events from consumer within a certain radius to the cluster centers when computing the popularity.
\Cref{fig:example_segment} shows the interaction history of example consumers that were identified to belong to one segment (left), the representative segment items identified in their segment (middle), and the fashion expert interpretation of the style (right).

\myparagraph{The Fashion Expert interpretation.} 
Our fashion expert assessed styles and assignments by evaluating consumer histories and representative segment items. 
We conducted this evaluation for segmentation using 20, 250, and 1000 clusters. As for the latter two, we sample around 20 segments based on size and average inter-cluster style similarity to make the evaluation feasible. We sampled 15 consumers from each of the selected clusters to be evaluated by a fashion expert.
The evaluation concluded that using 20 segments, there was some shared behavior for the consumers closest to the centroid. However, the more distant consumers in the same clusters exhibited weak similarity to consumers close to the centroid, indicating that the segments might be too large. 
Using 1000 segments, we observe that the segments are over-focused on a single item that is repeated many times throughout histories (e.g., a certain Nike sneaker or a leather belt). The segments in this case seemed to be too specific to carry a notion of style. 
Of the chose variants, the 250 segments showed the overall best alignment with a notion of style and consistency of consumers.
This aligns with our previous investigation on the center-distance distribution.

\myparagraph{Recommendations with data-driven segmentation}
Using the representative segment items, we proposed three approaches for style-segment based item recommendation:
\emph{Approach 1} replaces the recommendations entirely with the representative segment items;
\emph{Approach 2} backfills a percentage of the consumer’s interaction history with representative segment items and uses the resulting sequences as input to the existing recommendation algorithms to generate recommendation candidates;
\emph{Approach 3} interleaves the resulting output candidates of the existing recommendation algorithms with the representative segment items.
To determine the most suitable approach for style-based item recommendations, we conducted offline evaluations to measure how effectively style-based recommendations can improve the attributes diversity and relevance (nDCG, overlap coefficient with historical consumers’ interactions) of items. For nDCG, we use historical consumer click data on the items recommended by existing recommendation algorithms. We additionally conducted qualitative visual inspection of example recommendations. The results indicate that Approach 2 with 20\% backfilling of representative segment items had the best performance -  increasing brand and commodity group diversity for less engaged consumers, a relatively smaller loss in nDCG, and superior adaptability to consumer context compared to alternative approaches. Visual evaluations of the product recommendation conducted by fashion experts also confirmed that Approach 2 seamlessly combines consumer preferences with representative segment items, resulting in a more diverse product range while still maintaining individual consumer preferences.

\subsection{Lookalike segmentation of designer consumers}

We explore lookalike segmentation in the context of \textit{designer consumers}. In this use case, we define \textit{core designer consumers} as consumers with a specified minimum amount of interactions with designer brands over the last year, indicating a high affinity towards those brands.
Our goal is to find consumer that would likely show a high affinity towards designer brands in the future, in order to present them more relevant recommendations.

Using this core segment definition, we aim to differentiate the historical behavior of core designer consumers from that of non-designers consumers and subsequently identify lookalike designer consumers. Negative consumers were randomly selected from the pool of non-designer consumers, excluding new consumers for whom the first activity on the platform was less than 7 days ago.

\myparagraph{Data.} We create a dataset composed of timestamped events (article-click and add-to-wishlist) sequences per consumer labeled with the designer status (core or negative). To create training input sequences for the model we randomly sample 100 consecutive events from the last 4 months of the consumers' histories. We augment the dataset to increase the ratio of positive examples by sampling up to 5 non-overlapping sequences from the core consumers' histories. To create input sequences for inference we take the last 100 events of all consumers not in the core designer set. Additionally, we utilize consumer specific features, such as sales-channel, age segment and gender preference.

\myparagraph{Lookalike model.} Using this dataset we train a sequential model based on the transformer architecture~\cite{vaswani2017attention}.
 We use the following SKU features: brand, season-code, silhouette-code, tag, material, designer-status, price and whether the brand is followed by the consumer. The features specific to the consumer (age-segment, gender preference, and sales-channel) are provided by the CLS token. For categorical features, we use a trainable embedding layer with dimension 64. Numerical features, i.e. price and event timestamp, are normalized to the unit interval and multiplied by an embedding layer, effectively scaling a trainable embedding vector. 
 For each token, we sum the embedding of each of the features contributing to it.
 We use a softmax layer for classification, which uses the encoded CLS token as input and outputs scores for both classes – designer and non-designer. The model is trained using a binary-cross entropy loss predicting the designer status of the consumer sequence. 

 Besides the model described above \emph{Variant 1}, we consider four other variants to further investigate the proposed model; \emph{Variant 2} uses additional data balancing based on class weights, in order equalize the weight of both designers, and non-designers; \emph{Variant 3} uses piece wise linear encoding~\cite{gorishniy2022embeddings} to represent the numerical features of events (price and timestamp); \emph{Variant 4} combines Variant 3 with data balancing as described in \emph{Variant 2};
\emph{Variant 5} omits the timestamp feature of the events, which essentially removes the order of the events.

 \myparagraph{Classification threshold optimization.}
 To identify lookalike consumers, we need to set the threshold parameter $\tau$, as defined in \cref{eq:lookalike-set}. This parameter is a trade-off between the similarity of the lookalikes to the core consumers and the number of lookalikes obtained from the model.
 We identify the \Ftwoscore{} as a suitable metric to determine the quality of a set threshold. To find the optimal $\tau$, we compute the \Ftwoscore{} for a range of threshold and determine their optimum.

\myparagraph{Offline Experiments.} We evaluate five model variants using precision, recall, average precision and \Ftwoscore{}, the latter of which is our primary evaluation metric. 
\Cref{tab:lookalike} shows offline evaluation results for random baseline and the five transformer variants based on a single training and evaluation run. 
Our model outperforms the random classifier on all the metrics which serves as a basic sanity check. Model Variant 5 shows best performance on the majority of the metrics, achieving \Ftwoscore{} of  0.439 (+5.3\% compared to Variant 3, +6\% compared to Variant 1) with the optimized threshold. Precision with optimized threshold is better in Variant 4, but increase in the recall makes up for it when computing the \Ftwoscore.

\begin{table}
  \caption{Offline evaluation results for random baseline and the transformer-based classifier on four dataset variants.}
  \label{tab:lookalike}
  \begin{tabular}{rcccc}
    \toprule
    Variant&\Ftwoscore&Precision&Recall&Average Precision\\
    \midrule
    Random&0.087&0.02&0.5&0.02\\
    1&0.414&0.195&0.576&0.263\\
    2&0.412&0.202&0.558&0.265\\
    3&0.417&0.204&0.565&0.27\\
    4&0.415&\textbf{0.216}&0.540&0.269\\
    5&\textbf{0.439}&0.208&\textbf{0.608}&\textbf{0.293}\\
    \bottomrule
  \end{tabular}
\end{table}

\begin{figure}[h]
  \centering
  \includegraphics[height=0.3\linewidth]{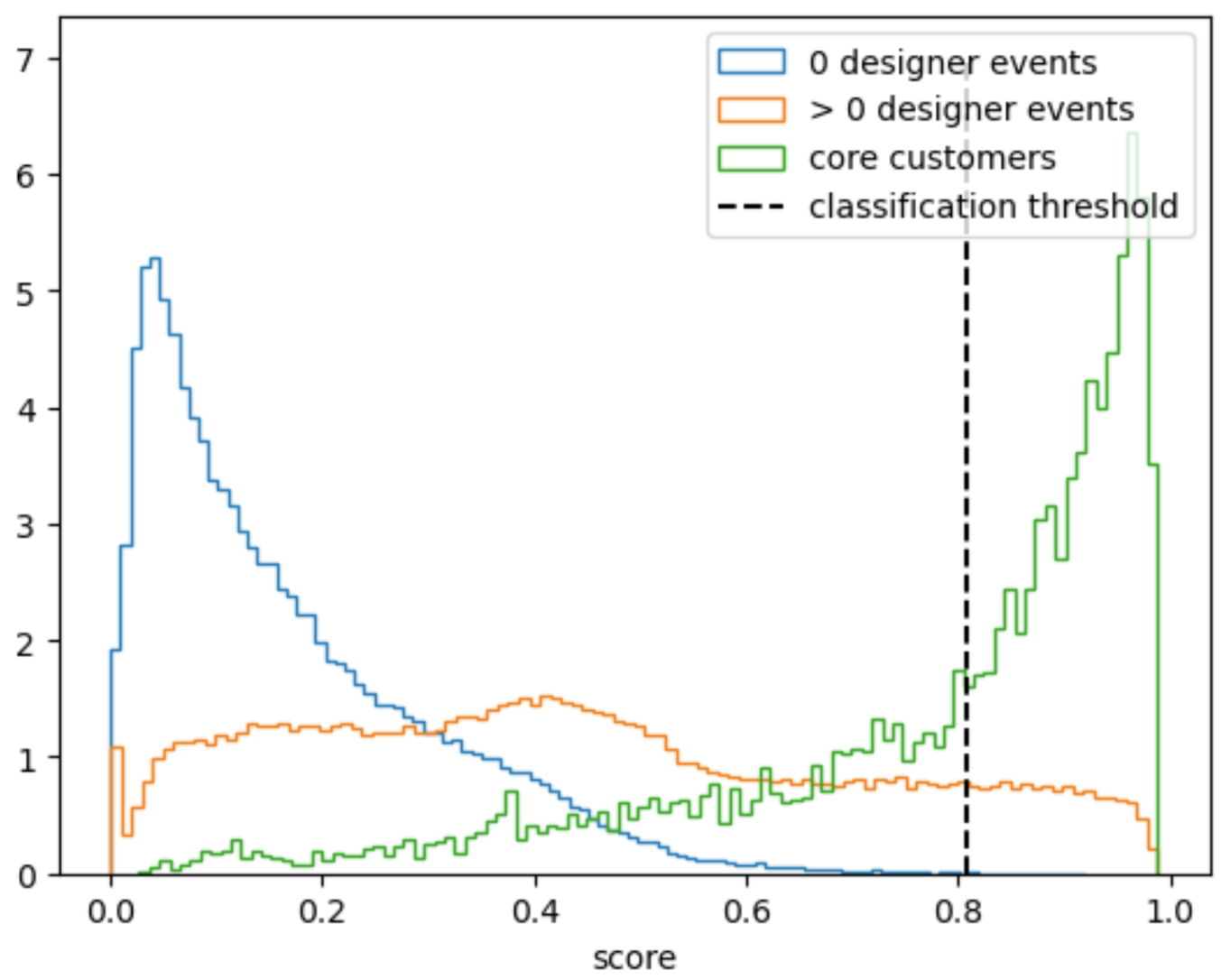}
  \includegraphics[height=0.3\linewidth]{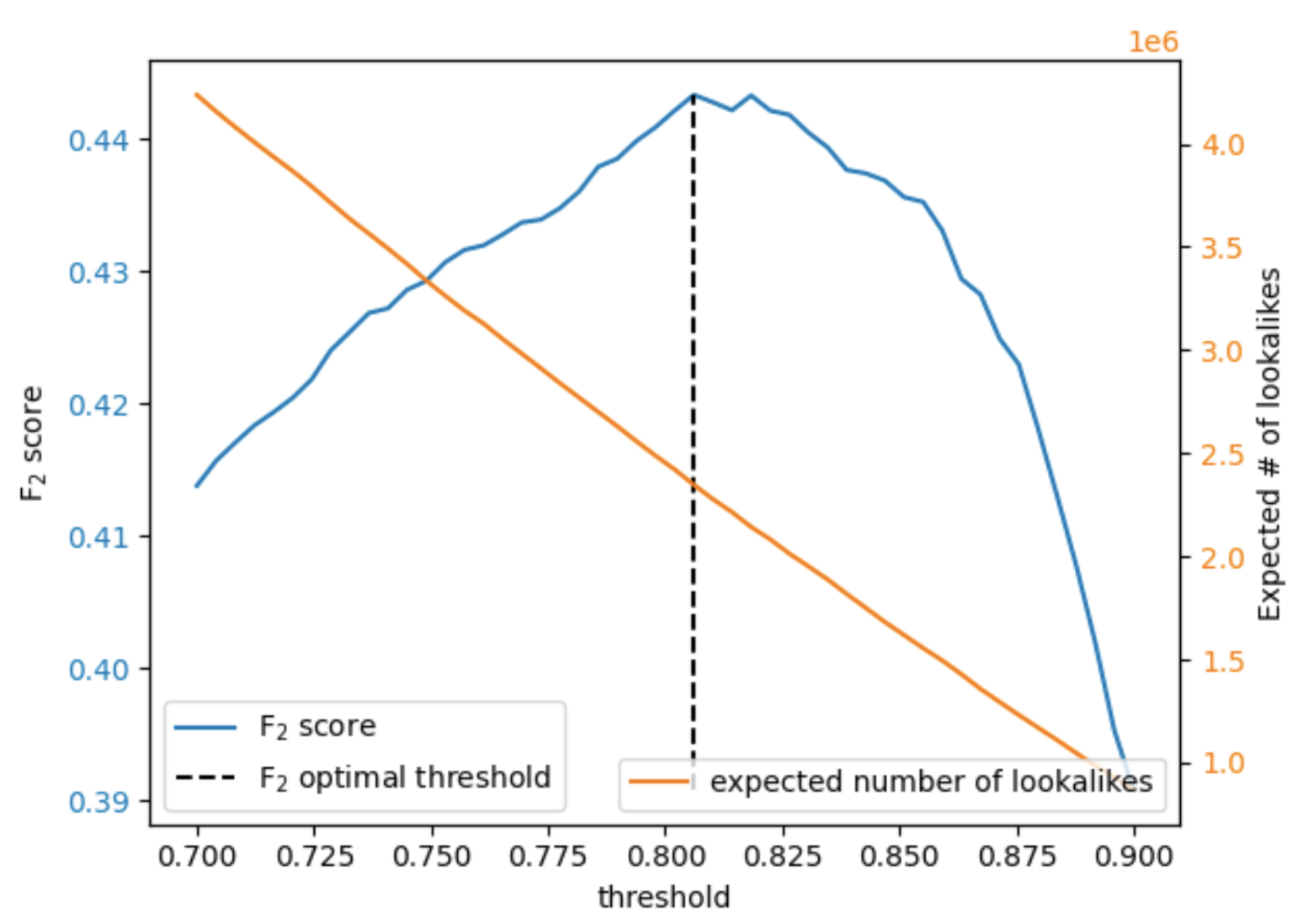}
  \caption{\textbf{Left:} Model predictions for core and non-core consumers on the validation set. Predictions on non-core consumers are further resolved by consumers with zero designer events  and consumers that have at least one designer events. The lookalike classification threshold (black dashed line) is determined by maximizing the \Ftwoscore{}. \textbf{Right:} \Ftwoscore{} and expected number of lookalike consumers depending on the threshold: By varying the classification threshold, we can vary the number of lookalikes. This curve allows for setting a threshold that might fulfill business needs, such as a minimum number of lookalikes, while still monitoring the quality of the lookalikes as indicated by the \Ftwoscore{}.}
  \Description{}
  \label{fig:preds}
\end{figure}

\Cref{fig:preds} (left) shows the distribution of scores for ground truth core and non-core consumers on the validation set. Non-core consumers are further divided into ones with zero designer interactions and with one or more designer interactions. These distributions indicate that the model is capable of solving the classification task separating core designer consumers from the rest. Furthermore, there is a large tail of non-core consumers whose score is in the core consumers distribution and all of these consumers have one or more designer interactions, indicating that these consumers already show some level of affinity to designer brands. These consumers are the designer lookalike consumers we are trying to extract. By varying the classification threshold we can extract a different number of lookalike consumers depending on the application. We cannot directly access the quality of these consumers, but we can calculate classification metrics on the validation set using the threshold and treating core consumers as ground truth positive examples. Conversely, we can first optimize the threshold on the validation set and then use it to extract the lookalike consumers.

\Cref{fig:preds} (right) shows the dependency of the \Ftwoscore{} and the number of lookalike consumers on the classification threshold. Increasing the threshold reduces the number of consumers classified as lookalikes. Relation with the \Ftwoscore{} is more complex, peaking at the threshold value of 0.808. This threshold value gives us 2.31M of designer lookalike consumers.

\myparagraph{Online experiments.}
We ran an online experiment to test the lookalike Designer consumers on the product catalog page. To this end, we equally split the set of lookalike consumers into a treatment and control group. The two-phased ranking model first fetches the available items along with their attributes and popularity scores and then ranks them based on the consumer preferences by considering their interaction history. For the control group, we use the default ranking model, whereas for the treatment group we use a variant fine-tuned on the \textit{core designer consumers} set which ranks items from designer brands higher in the catalog.
The expectation was that by offering a ranking experience tailored for designer, the lookalike consumers would be able to discover more items that resonate with their preferences and their aspiration to become Designer consumers compared to the default general ranker approach. The A/B test ran for around 5 weeks and reached ~1.64M lookalike designer consumers. Online KPIs primarily relied on the performance of the ranking system and did not directly measure the performance of the lookalike model itself. As a result, we detected a significant positive increase click through rate (0.45\%).

\section{CONCLUSION AND NEXT STEPS}
\label{sec:conclusion}
In this work, we presented UNICON - a unified consumer segmentation framework capable of driving personalization in fashion e-commerce. Our approach consists of learning dense long-term consumer representations that are then utilized to derive two essential classes of consumer groups: lookalike and data-driven.  We demonstrated the capability of this framework using real-world large-scale consumer data to identify \textit{lookalike designer segment} - consumers with high affinity towards premium fashion brands - and \textit{data-driven style groups} - segments of consumers exhibiting long-term similar fashion style. Through offline and online tests, we showed that the identified lookalike consumers display high engagement with designer items and that we have improved their experience. Additionally, our comprehensive experimentation with style segments showed the gains of employing consumer groups in a hybrid personalized recommender system that complements strong individual preferences with wider group interests. Future work consists of investigating methods to improve the consumer embeddings to reflect more robust and longer-term behavior representation by expanding the sequence length in the model, extending the  considered interactions time frame, and exploring alternative approaches to summarize the consumer embedding other than simple averaging over the embedded token sequence. For lookalike modeling, we want to compare our model with more straightforward baseline models. Additionally, for the unsupervised clustering of data-driven groups, experimenting with more sophisticated approaches such as combining dimensionality reduction using UMAP~\cite{mcinnes2018umap} with density based clustering using HDBSCAN~\cite{mcinnes2017hdbscan} would reduce the impact of outlier consumers on the formation of the clusters and enable automatic selection of the number of groups.

\section{ACKNOWLEDGMENT}
We thank Ellen Scherer for her valuable insights in interpreting fashion style, Mathilde Caron for driving the product vision of this project, and both Marjan Celikik and Evertjan Peer for their contributions in the early stages of this line of work.  
\newpage
% \printbibliography
\bibliographystyle{ACM-Reference-Format}
\bibliography{main}
\end{document}